\documentstyle[amssymb,preprint,aps]{revtex}
\tightenlines
\begin{document}
\draft
\title{
Combinatorial Optimization by Iterative Partial Transcription
}
\author{A.~M\"obius$^1$\footnote{e-mail: a.moebius@ifw-dresden.de}, 
B.~Freisleben$^2$, P.~Merz$^2$, and M.~Schreiber$^3$  \\  }
\address{
$^1$Institut f\"ur Festk\"orper- und Werkstofforschung,
D-01171 Dresden, Germany,\\
$^2$Fachbereich Elektrotechnik und Informatik, 
Universit\"at-Gesamthochschule,
D-57068 Siegen, Germany,\\
$^3$Institut f\"ur Physik, Technische Universit\"at, D-09107 Chemnitz, 
Germany
}
\date{\today}
\maketitle
\begin{abstract}
A procedure is presented which considerably improves the performance 
of local search based heuristic algorithms for combinatorial 
optimization problems. It increases the average ``gain'' of the 
individual local searches by merging pairs of solutions: certain parts 
of either solution are transcribed by the related parts of the 
respective other solution, corresponding to flipping clusters of a 
spin glass. This iterative partial transcription acts as a local 
search in the subspace spanned by the differing components of both 
solutions. Embedding it in the simple multi-start-local-search 
algorithm and in the thermal-cycling method, we demonstrate its 
effectiveness for several instances of the traveling salesman problem. 
The obtained results indicate that, for this task, such approaches are 
far superior to simulated annealing.
\end{abstract}



\section{Introduction}

Combinatorial optimization problems occur in many fields of physics, 
engineering and economics. They are closely related to statistical 
physics, see e.g.\ \cite{Usam.Kita,Zeng.etal} and references therein. 
Many of the combinatorial optimization problems are difficult to solve 
since they are NP-hard, i.e., there is no algorithm known which finds 
the exact solution with an effort proportional to any power of the 
problem size. One of the most popular such tasks is the traveling 
salesman problem (TSP): how to find the shortest roundtrip through a 
given set of cities. For recent surveys on various approaches to the 
TSP see \cite{John.McGe,Jung.etal}. 

Many combinatorial optimization problems are of considerable practical 
importance. Thus, algorithms are needed which yield good 
approximations of the exact solution within a reasonable computing 
time, and which require only a modest effort in programming. Various 
deterministic and probabilistic approaches, so-called search 
heuristics, have been proposed to construct such approximation 
algorithms. A considerable part of them borrows ideas from physics and 
biology.

The conceptionally simplest approximation algorithms are {\em local 
search} procedures. They can be best understood when interpreting the 
approximate solutions as discrete points (states) in a 
high-dimensional hilly landscape, and the quantity to be optimized as 
the corresponding potential energy. These algorithms proceed 
iteratively, improving the solution by small modifications (moves) 
step by step: The neighborhood of the current state, defined by the 
set of permitted modificiations of the solution (move class), is 
searched for states of lower energy. If such a state is found, it is 
substituted for the current state, and a new search is started. 
Otherwise, the process stops because a local minimum has been reached. 

Usually, the chances to find the global minimum in this way -- or in 
the case of multimodality (degeneracy), one of the global minima --
vanish exponentially as the problem size rises. They can be increased 
by taking moves of higher complexity into account. Physically 
speaking, by means of the local search we create states which are only 
metastable; the degree of metastability is defined by the move class 
considered. Thus, according to increasing complexity of the moves, one 
can define hierarchies of classes of metastable states. Considering 
more complex moves corresponds to waiting longer relaxation times, so 
that, on the average, one ends up in lower local minima.

The local search concept is simple. However, in sophisticated 
algorithms, the moves considered can be fairly complicated, i.e., they 
may concern a rather large number of degrees of freedom as in the 
Lin-Kernighan algorithm for the TSP \cite{Lin.Kern}. The art of 
developing such algorithms is to select from the set of all possible 
modifications, related to a given number of degrees of freedom, an 
appropriate small part to be included into the move class.

In order to overcome barriers between local minima, {\em simulated
annealing} (SA) \cite{Kirk.etal,Laar.Aart} assumes the ``sample''
(current approximate solution) to be in contact with a heat bath with
a time dependent temperature. Thus, moves increasing the energy are
also taken into account, where the acceptance probability decreases 
exponentially with increasing energy change. Slow cooling permits the 
``sample'' to reach a particularly deep local minimum. 

Several proposals have been made to improve this basic concept, in
particular to optimize the temperature schedule of the annealing 
process, see e.g.\  
\onlinecite{Szu.Hart,haj88:cso,Ingb.89,Ingb.96,bt96:osa,Mari} and
references therein, or to adapt SA to parallel computer architectures
\cite{soh95:pns,nz95:psa,rss96:psa,Schn.etal}. Moreover, substituting 
the random decision of accepting energy increasing moves by
a deterministic decision according to whether or not the energy
change exceeds a certain upper bound, one gets {\em threshold
accepting}, a closely related concept \cite{ds90:ta}. Finally, 
replacing the slow cooling of SA by {\em thermal cycling}, i.e., by 
cyclically heating and quenching with decreasing amplitude, can 
considerably improve the performance \cite{thercycl}; for an early 
approach based on cyclically heating (with the temperature chosen at 
random) and rapid cooling see \cite{MuKr}.

{\em Genetic algorithms} \cite{hol75:ana,gol89:ga} offer another 
possibility to escape from local minima. They simulate a biological 
evolution process by operating on a population of individuals 
(approximate solutions), where new generations are produced through 
the repeated application of genetic operators such as selection, 
crossover and mutation. Particularly effective seem to be algorithms 
in which the individuals are local minima, see 
\cite{Brad,Muhl.etal,Frei.Merz.a,Frei.Merz,Merz.Frei} and references 
therein. However, a guarantee to find the global optimum within any 
finite computing time cannot be given by this approach, nor by any 
other of the heuristic methods mentioned, though, for infinite 
computing time, the convergence of SA (with a logarithmic temperature 
schedule) and of a broad class of evolutionary algorithms has been 
proved \cite{Gema.Gema,Rudo}.

At the same time, exact solution methods have been developed further. 
They are mainly based on branch-and-bound and branch-and-cut ideas
\cite{Bala.Toth,Jung.etal2}. Thus, a specific TSP instance including 
7397 cities was solved \cite{John.McGe,Rein}. However, for a fixed 
size, the effort necessary to find the exact solution can vary 
enormously from problem to problem. For example, the TSPLIB95 
\cite{Rein} includes an instance of 1577 cities which could only
very recently be solved by Applegate and co-workers; they needed 
approximately 280 hours on a DEC Alphastation 4100 5/400 \cite{Apple}.

In this paper, we present {\em iterative partial transcription} (IPT), 
an approach to improve the performance of heuristic algorithms for 
combinatorial optimization problems: IPT compares pairs of states, 
represented by vectors of coordinates. In an iterative procedure, it 
systematically searches for the subsets of the components of these 
vectors, the copying of which from one vector to the other yields new 
approximate solutions with decreased energy. IPT is particularly 
useful when applied to local minima. We illustrate its efficiency
for the TSP, demonstrating that the incorporation of IPT into local 
search based heuristic algorithms can considerably increase their 
performance.

The paper is organized as follows: In Section II, we present the
IPT procedure in a general manner, as well as applied to the TSP. 
Section III is devoted to embedding IPT in multi-start local search 
and in thermal cycling. Section IV reports on the results obtained for 
several instances of the traveling salesman problem. Finally, Section 
V summarizes the paper.

\section{Iterative partial transcription}

\subsection{General formulation}

The design of the optimization method proposed here is motivated by
the use of local search algorithms common to three highly effective 
Monte Carlo optimization procedures: the {\em iterated Lin-Kernighan} 
method for the TSP \cite{John.McGe,Mart.etal,Mart.Otto}, the 
{\em thermal-cycling} approach \cite{thercycl}, and the 
{\em genetic-local-search} strategy 
\cite{Muhl.etal,Frei.Merz.a,Frei.Merz,Merz.Frei}. The efficiency of 
these approaches in finding states of particularly low energy rests on 
the consideration of local minima rather than of arbitrary states, and 
on modifying the local minima by sophisticated operations. These 
operations typically involve elaborate manipulation steps, and in some 
cases they make use of other local minima. Compared to SA, a single 
such modification concerns a rather large number of degrees of 
freedom. Of course, it demands far more CPU time than a single 
Metropolis step in SA. The idea of our proposal is to increase the 
average ``gain'' of the individual local searches by a fast 
postprocessing phase, and consequently to reduce the average number of 
local search steps required to reach a certain energy. This is 
achieved by making good use of the information inherent in the 
transformation mapping one to the other local minimum.

In detail, consider two states ${\bf v}_1$ and ${\bf v}_2$ (possible 
approximate solutions of the given optimization problem), encoded as 
two vectors, which differ in $k$ components. We look for 
decompositions of the transformation ${\bf M}$ mapping ${\bf v}_1$ to 
${\bf v}_2$ into a product of two commuting transformations 
${\bf M}_\alpha$ and ${\bf M}_\beta$, 
\begin{equation}
{\bf v}_2 = {\bf M}({\bf v}_1) 
= {\bf M}_\beta({\bf M}_\alpha({\bf v}_1)) 
= {\bf M}_\alpha({\bf M}_\beta({\bf v}_1))
\end{equation}
such that ${\bf M}_\alpha({\bf v}_1)$ and ${\bf M}_\beta({\bf v}_1)$ 
are possible approximate solutions of the optimization problem too, 
and that ${\bf M}_\alpha$ and ${\bf M}_\beta$ modify disjunct sets of 
$k_\alpha$ and $k_\beta$ components of ${\bf v}_1$. Thus, 
$k_\alpha + k_\beta = k$, so that both ${\bf M}_\alpha$ and 
${\bf M}_\beta$ ``transcribe'' part of the components of ${\bf v}_1$ 
by the values of these components in ${\bf v}_2$. 

The procedure proposed here is an iterative search for appropriate 
transformations of this kind. It merges two states ${\bf v}_1$ and 
${\bf v}_2$: According to increasing $k_\alpha$, where 
$1 \le k_\alpha \le k$, it systematically searches for pairs of 
${\bf M}_\alpha$ and ${\bf M}_\beta$ satisfying Eq.\ (1). If such a 
pair is found, it checks whether or not ${\bf M}_\alpha$ improves 
${\bf v}_1$. If yes, ${\bf v}_1$ is substituted by 
${\bf M}_\alpha({\bf v}_1)$, otherwise ${\bf v}_2$ by 
${\bf M}_\alpha^{-1}({\bf v}_2) = {\bf M}_\beta({\bf v}_1)$, and 
then the search is restarted. The iteration stops if 
${\bf v}_1 = {\bf v}_2$. This procedure, which we refer to as {\em 
iterative partial transcription} (IPT), has as its output the current 
${\bf v}_1$.

Below, we apply IPT to local minima with respect to some move class. 
However, the IPT output state will in general not be such a local 
minimum. Therefore, provided the IPT output state differs from both 
the input states, it is additionally exposed to a local search with 
respect to this move class. We refer to this combination of IPT and 
local search as IPTLS.

The proposed IPT procedure decomposes the rather complex 
transformation of one state to another into several parts, analyzing 
with respect to which features these states differ. Disregarding the 
disadvantageous features, it effectively makes use of the favorable 
ones for a specific improvement. This approach can easily be 
understood when it is interpreted in physical terms: We consider 
low-energy states as differing from the ground state by several 
non-interacting ``elementary'' excitations, which, however, may 
involve rather complex modifications. Comparing two low-energy states, 
we identify the excitations which are present in one of these states, 
but not in the other, and generate a new low-energy state by 
relaxation of all the excitations found. In this sense, IPT is a 
generalization of the basic idea of the approach to finding the ground 
state of a spin glass proposed by Kawashima and Suzuki 
\cite{Kawa.Suzu}. These authors relax excitations formed by clusters
of neighboring spins, which they identify by the comparison of 
different replicas.

There are some links between this method and other heuristic search 
algorithms: IPT can be considered as a local search in the subspace 
spanned by the differing components of both states. The related move 
class is given by the possibilities of simply inheriting a ``part'' of 
the other state, which corresponds to a shift to the alternative point 
in a particular subspace of the configuration space. As the 
Lin-Kernighan procedure for the TSP \cite{Lin.Kern}, IPT takes rather 
complex moves into account while diminishing the effort needed for 
exploring the search space by largely reducing its dimension. 
Alternatively, in biological terms, IPT can be interpreted as the 
deterministic transcription of (groups of) genes. 

IPT is applicable to several problems. For example, for the TSP,
${\bf M}_\alpha$ would correspond to the transcription of a part of
the tour; for a short-range Ising spin glass, ${\bf M}_\alpha$ would 
describe the flipping of a cluster of neighboring spins, cf.\ Ref.\ 
\cite{Kawa.Suzu}. However, IPT is clearly not applicable to problems 
with long-range interaction such as the Coulomb glass (an Ising spin 
glass with Coulomb interaction).

\subsection{Realization for the TSP}
 
We illustrate IPT by applying it to the traveling salesman problem. 
The states (possible solutions) are permutations of the $N$ given 
cities. The length of the roundtrip corresponds to the potential 
energy to be minimized. We use the following notions: tour and subtour 
denote closed roundtrips through all cities and part of the cities, 
respectively, whereas chains and subchains stand for tours and 
subtours with one connection eliminated, respectively. The number of 
cities in a subchain is referred to as its size. Thus, to identify 
pairs of transformations ${\bf M}_\alpha$ and ${\bf M}_\beta$ in the 
sense of the general description of IPT means to search for subchains 
which include the same cities in a different order, and have the same 
initial and final cities. 

Starting from two tours A and B, IPT proceeds according to the 
following scheme:
\begin{itemize}
\item[(1)] 
Formation of a reduced representation: For each city, check whether 
or not it has the same neighbors in both tours / subtours. If yes, 
create a new pair of subtours by omitting this city and connecting its 
neighbors. Let the number of cities in the reduced problem be 
$N_{\rm r}$. The ``next'' cities of $i$ in $A$, i.e., the cities
following the city $i$ in tour $A$ of the reduced problem, are denoted 
by $n^A_{i,1}$, $n^A_{i,2}$, $n^A_{i,3}$, and so on; the ``previous'' 
cities of $i$ are named $p^A_{i,1}$, $p^A_{i,2}$, $p^A_{i,3}$, and so 
on. The cities of tour $B$ are referred to analogously.
\item[(2)] 
Comparison of subchains of the reduced tours $A$ and $B$ where their 
size $s$ increases from 4 to $N_{\rm r}/2+1$: Check for all $i$, 
whether the final cities are the same, that is, whether 
$n^A_{i,s-1} = n^B_{i,s-1}$, or alternatively 
$p^A_{i,s-1} = n^B_{i,s-1}$. Provided one of these conditions is 
fulfilled, investigate whether or not the corresponding subchains 
include the same cities \cite{touch}. If yes, substitute in the 
original tours the worse of the corresponding subchains by the better 
one (in the case of equality, substitute the corresponding subchain 
in $B$), and go to (1).
\item[(3)] 
Choose the better of the current original tours $A$ and $B$ to be the 
IPT output.
\end{itemize} 

Our IPT algorithm for the TSP has some resemblance to the subroute 
transcription procedure originally proposed by Brady \cite{Brad},
later adopted by Yamamura et al.\ \cite{SXX,Maekawa} in the ``subtour 
exchange crossover'' operator of a genetic TSP algorithm. However, 
these two methods do not require to fulfill the restriction that the
two subchains must have the same initial and final cities. This 
condition is substantial in our approach: It guarantees that each 
transcription of a subchain diminishes the tour length. Moreover, it 
largely reduces the number of pairs of subchains to be compared in 
detail (whether or not they include the same cities), and thus the CPU 
time as well.

\section{Main algorithm}

The effectiveness of the IPT procedure can only be judged in the 
context of the main algorithm in which it is embedded. As such, we 
consider the multi-start-local-search algorithm and the 
thermal-cycling algorithm. In both cases, IPT acts on local minima 
only. Thus, we always use it in combination with an additional local 
search on output states differing from both the input states, that is 
in the IPTLS version.

\subsection{Multi-start local search}

The simplest manner of using a multi-start-local-search algorithm for 
the solution of an optimization problem is to perform $K$ times a 
local search starting from a random state, and to take the lowest of 
the resulting states as the final state. This algorithm is primitive, 
but it has the advantage of having only a single adjustable parameter, 
$K$. 

Incorporating IPTLS into this multi-start local search permits to 
combine the information obtained by the individual trials more 
efficiently. For that, the first approximation of the solution is 
obtained by a local search starting from a random state. Then, for 
$j=2$ to $K$, IPTLS is performed between the $(j-1)$-th approximation 
and the state obtained by the $j$-th local search starting from a 
random state. The output state is considered as the $j$-th 
approximation.

The performance of this extended multi-start-local-search approach is 
likely to improve when ``searching in parallel'', cf.\ 
\cite{Brad,Hube.etal}. In order to do so, we utilize an archive of 
$N_{\rm a}$ states ($N_{\rm a} < K$), where the state of lowest energy 
is considered as the current approximation. The archive is initialized 
by $N_{\rm a}$ local searches starting from states chosen at random. 
After this, $(K - N_{\rm a})$ times the following steps are performed: 
A new state is generated by a local search starting from a random 
state. Then, a series of IPTLSs is performed between this new state 
and the archive states. As soon as the resulting state has a shorter 
tour length than the currently selected archive state, it is 
substituted for this archive state, and the series of IPTLSs is 
terminated. Finally, after finishing these $K$ local searches extended 
by IPTLS, we try to improve the archive by applying IPTLS to all pairs 
of states contained in it. 

The ``searching in parallel'' approach is promising for three reasons: 
This method is, in effect, a partition of the computational effort 
into several search processes in order to minimize the failure risk 
\cite{Hube.etal}. More importantly, the low-energy states, created 
during the expensive local search starting from random states, are used 
multiply by means of the series of IPTLSs. Finally, the local search 
step following a ``successful'' IPT has to be performed at most once 
within each series.

\newpage

\subsection{Thermal cycling}

Thermal cycling \cite{thercycl} has been shown to be far more 
efficient than multi-start local search. It consists of cyclic 
heatings and quenchings by Metropolis and local search procedures, 
respectively, where the amount of energy deposited into the sample 
during the individual heatings decreases in the turn of the 
optimization process. This algorithm works particularly well when 
applied to an archive of $N_{\rm a}$ samples rather than to a single 
sample. 

The embedding of IPTLS in thermal cycling is achieved in the 
following three ways:
\begin{itemize}
\item[(i)]
The multi-start local search creating the initial archive is enhanced 
by additional IPTLS as described in the previous subsection. 
\item[(ii)]
Each temperature step starts with trying to improve the archive by 
applying IPTLS to all pairs of archive states, where the output state 
always replaces the better of the two input states -- substituting the 
worse of the two would cause a too early loss of variety in the 
archive, cf.\ \cite{Brad}.
\item[(iii)]
After each thermal cycle, a series of IPTLS between the final state 
and all archive states with energies smaller or equal to that of the 
initial state is performed. This series is terminated as soon as one
of the archive states is improved by the corresponding IPTLS step. In 
this sense, each thermal cycle does not act on its initial state only, 
but on (a part of) the whole archive.
\end{itemize}

Moreover, the inclusion of IPTLS between the final state of each cycle 
and the archive states suggests a change in the heating process. In 
Ref.\ \cite{thercycl}, a constant number of modifications is performed 
for heating, independent of the problem size. Now, this number is 
chosen to be proportional to the problem size. The reason for this 
change is the following. For very large problems, the total 
modification of the state within one heating-quenching cycle should 
frequently be a superposition of independent, ``local'' variations. 
Most of these variations cause an increase of the energy. Thus, in 
\cite{thercycl}, their number must be small to have a realistic chance 
for a net improvement. However, when IPTLS is included for 
postprocessing, the undesirable ``local'' variations are filtered out 
to a large extent, such that the above restriction can be abandoned.

\section{Application tests}

\subsection{Implementation details}

We now demonstrate the efficiency of the two algorithms described in
Sections III.A and III.B, respectively, for the TSP. These algorithms 
rely on an adequate local search procedure. Here, we use a slightly
improved version of the local search implementation of Ref.\ 
\cite{thercycl}. Thus, we have the choice between four alternative 
possibilities concerning the kind of metastability to be reached:
\begin{itemize}
\item[(a)] stable with respect to reverse of a subchain, as well as to 
shift of a city;
\item[(b)] same as (a), and stable with respect to cutting three 
connections of the tour, and concatenating the three subchains in a 
new manner;
\item[(c)] same as (b), and stable with respect to rearrangements by 
first cutting the tour twice and forming two separated subtours, and 
connecting then these subtours after cutting two other connections;
\item[(d)] same as (c), and stable concerning a restricted
Lin-Kernighan search \cite{Lin.Kern} which consists of cutting the
tour once, then several times alternately cutting the chain and
concatenating the subchains, and finally connecting the ends of the
chain again, where the number of trials to modify the chain is
restricted to 1000.
\end{itemize}
In the present study, we have performed numerical experiments
considering move class (a) or (d) mainly.

The efficiency of our local search approach rests on three principles:
(i) New connections are tried according to increasing length, where 
appropriate bounds are utilized to terminate the search as soon as it 
becomes useless. (ii) In stage (c), we first tabulate all 
rearrangements, which decompose the original tour into two subtours 
with a shorter total length. Then, we search for those decompositions 
of the original tour into two subtours, starting from which one of
the tabulated rearrangements produces a new, shorter tour. (iii) 
Limiting the number of trials in (d) improves the efficiency 
considerably if the cities are clustered, i.e., if a few of the 
distances between neighboring cities in the optimal tour are much 
larger than the others.

The IPT part is the same in both of the presented algorithms. It 
requires a computational effort which is roughly proportional to 
$N^2$. However, due to the use of a reduced representation, the 
proportionality constant seems to be small in practice: We performed 
multi-start-local-search-with-IPTLS runs ($N_{\rm a} = 1$) for sets of 
cities, randomly distributed in a square, with Euclidian metric. We 
observed that, for up to several thousand cities, even when only move 
class (a) is taken into account, the CPU time for the IPT is roughly 
one order of magnitude smaller than the CPU time for the local search.

Our thermal-cycling code \cite{thercycl} was adapted to using IPTLS in 
three points: (i) Since IPTLS ensures a high quality of the primary 
archive, the corresponding effort could be diminished; we now perform 
$30 \, N_{\rm a}$ rather than $50 \, N_{\rm a}$ searches starting from 
random states in initializing the archive. (ii) The heating part in 
thermal cycling, see Section III.B, has been changed in comparison to 
\cite{thercycl} according to the last paragraph of the previous 
section; each heating is terminated after $N/10$ rather than after 50 
modifications of the tour. (iii) Due to the efficiency enhancement of 
the individual thermal cycles by IPTLS, we now perform 
$2 \, N_{\rm a}$ rather than $5 \, N_{\rm a}$ cycles before deciding 
whether or not the temperature can be decreased. All other adjustable 
parameters of thermal cycling have the same values as in Ref.\ 
\cite{thercycl}.

For comparison, we have also performed a series of runs of a carefully 
tuned SA code, where the adjustable parameters were optimized for the
instance considered. In this or that way we took into account all the 
essential points discussed in the simulated annealing section of the 
TSP review \onlinecite{John.McGe}. Our program uses an adaptive 
temperature schedule, and automatically shrinks the move class 
utilized in the turn of the cooling process. 

More specifically, as starting temperature of SA, we choose $1/10$ of 
the length reduction when quenching a random tour, divided by the number 
of cities. At each temperature, we perform a given number of sweeps. 
Then, if during this series of sweeps the best state found so far could 
not be improved, we decrease the temperature by a factor 0.9; otherwise 
we perform the same number of sweeps with unchanged temperature again, 
and so on. Finally, after 10 temperature steps without improvement of 
the best state so far, we terminate the cooling, and, for this best 
state, we perform a local search considering the complete move class 
(a). This adaptive exponential schedule is robust concerning moderate 
changes of the initial temperature. In optimizing our implementation, we 
have also tried logarithmic and $1/k$ schedules. But none of them lead 
to a clear acceleration compared to the schedule described. 

In our implementation, we construct the SA move class starting from 
the local-search move class (a), and restricting it by neighborhood 
pruning. This  means that the number of neighbors considered in 
selecting the first of the new connections of a move \cite{deta} is 
temperature dependent: We choose the upper bound of the corresponding 
neighbor identification number (1 for nearest neighbor, 2 for 
next-nearest neighbor, and so on) as 2.5 times its mean value for the 
tour modifications performed within the previous series of sweeps. This 
neighborhood pruning is very effective; without it, the program would 
be slower by roughly a factor of 40 (for 1~$\%$ accuracy).

The numerical experiments reported in this paper were performed using
an HP K460 with 180 MHz PA8000 processors, running under HP-UX 10.20.
(All CPU times given relate to one processor.) Our code was written in 
FORTRAN77.

\subsection{Multi-start-local-search results}

Since heuristic procedures yield only approximate solutions, the truly 
important property is the relation between the mean quality of the 
solution, that is the deviation of the mean tour length from the 
global optimum, and the required computing time, $\tau_{\rm {CPU}}$. 
Thus, in order to illustrate the performance of IPT, we have 
investigated the influence of the adjustable parameters on this 
relation for the 532 Northamerican cities problem (att532), a standard 
example from the TSPLIB95 \cite{Rein}. These results are presented in 
Figs.\ 1 to 3. Moreover, to check for robustness and size dependence, 
we have additionally studied five other instances from the TSPLIB95, 
i.e., pcb442, rat783, fl1577, pr2392, and fl3795, considering a 
smaller number of parameter sets, see Tables I and II. (Except for
pr2392, the instances chosen are the same as in \cite{thercycl}.)

The performance of multi-start local searches with move classes (a) 
and (d), respectively, is shown in Fig.\ 1 for att532. This graph 
contrasts results obtained for $N_{\rm a} = 1$ with and without 
IPTLS, and includes SA data (cf.\ previous subsection) for comparison. 
In particular, Fig.\ 1 shows the following:
\begin{itemize}
\item[(i)]
For large $\tau_{\rm {CPU}}$, i.e., for a large number of local 
searches $K$, considerable performance gains are reached when the 
multi-start local search is extended by IPTLS. This is observed for 
move class (a), as well as for move class (d). The speed gains are 
small when $K$ is close to 1, but they rapidly increase with $K$. For 
the highest $K$ considered, they amount to factors of roughly 100 and 
30 for multi-start local searches concerning (a) and (d), 
respectively. In further experiments, we obtained analogous results 
for move classes (b) and (c). 
\item[(ii)]
Even without IPTLS, multi-start local search using a sufficiently 
complex move class can be clearly advantageous in comparison to SA 
\cite{Moeb.Rich}: compare the multi-start-local-search data for move 
class (d) with the SA results. 
\item[(iii)]
For att532, if an accuracy between 1 and 3~$\%$ is required, even 
multi-start local search according to move class (a) extended by IPTLS 
can compete with SA: it is a bit better for 
$\tau_{\rm {CPU}} < 4\ {\rm sec}$, and slightly worse for larger 
$\tau_{\rm {CPU}}$. However, if higher accuracies are desired, 
our SA program outperforms the multi-start-local-search-with-IPTLS code, 
which utilizes only move class (a). A minor result of this comparison, 
not obvious from the figure since each point represents an average of 
100 runs, concerns the variance of the final tour length: the variance 
is considerably smaller for the multi-start local search according to 
(a) extended by IPTLS than for SA.
\end{itemize}

The advantage of ``searching in parallel'' \cite{Brad,Hube.etal} is 
demonstrated by Fig.\ 2. We compare the 
multi-start-local-search-with-IPTLS results of Fig.\ 1 to data 
obtained with archives of 3 and 10 states. For small numbers of local
searches $K$, there is almost no influence of ``parallelizing'', i.e.,
of using $N_{\rm a} > 1$. However, as $K$ increases, corresponding to 
increasing $\tau_{\rm CPU}$, the ``searching in parallel'' strategy
performs better and better. Moreover, up to some optimum archive size, 
the advantage increases also with $N_{\rm a}$. Above the optimum size,
the performance slightly decreases with increasing $N_{\rm a}$: 
additional runs for move class (d) showed that, in the whole accuracy 
region presented in Fig.\ 2, the performance decreases a bit when the 
archive size increases from 10 to 30. The optimum archive size seems 
to rise slowly with $K$.

The SA data, given in Fig.\ 1, are included into Fig.\ 2 also. It is 
remarkable that for att532 multi-start local search with IPTLS 
performed in parallel ($N_{\rm a} = 10$) has roughly the same 
performance as our tuned SA program. However, the former method has the 
considerable advantage to possess only one tuning parameter, which is, 
moreover, rather ``uncritical''.

To ensure fairness of the comparison, we have implemented the 
searching-in-parallel idea also in our SA program: four runs, each 
taking one fourth of the available computing time, are performed, and 
the best tour found in these runs is taken as final result, cf.\ 
\onlinecite{John.McGe,Brad,Hube.etal}. This performance curve is 
presented in Fig.\ 2. There is a clear efficiency increase arising 
from this parallelism if an accuracy better than 1~$\%$ is desired. 
However, for att532, as Fig.\ 2 shows, even this sophisticated SA 
algorithm is still far slower than the multi-start local search with 
IPTLS concerning move class (d). 

For a broader test of our code, we considered six symmetric TSP 
instances taken from the TSPLIB95 \cite{Rein}, including between 442 
and 3795 cities. Table I presents results for three different parameter
sets of trials, $K$, and archive sizes, $N_{\rm a}$. These data 
confirm the above interpretations concerning the performance of our 
algorithm: 
\begin{itemize}
\item[(i)] 
For all instances considered but pr2392, the best known tour lengths 
\cite{Rein} were reproduced. For pcb442, att532, rat783, and fl1577, 
these values are the exact optima; fl3795 has not been solved exactly 
yet. For pr2392, our best (mean) result exceeds the known exact 
optimum tour length by 0.5~$\%$ (1~$\%$).
\item[(ii)] 
There is a considerable benefit of ``searching in parallel'' as 
illustrated by the results for $K = 500$ with archive sizes 1 and 10, 
respectively.
\end{itemize}

Finally, the comparison of the data in Table I with those given in 
Table I of \cite{thercycl} is instructive. However, this consideration
is complicated by the use of a slightly improved local search code 
in the present work, which typically causes a speed gain by a factor
of 1.5. The comparison shows that multi-start local search extended 
by IPTLS and performed ``in parallel'' reaches roughly the same 
efficiency as thermal cycling without IPTLS. In more detail, the 
former program is clearly faster for att532, fl1577 and fl3795, but 
slower for rat783. For pcb442, both codes have roughly the same 
performance. The fact that there is no clear size dependence in this 
comparison is not surprising due to the large variety of the features 
(occurence of clusters of cities, degeneracies, ...) of the examples 
considered.

\subsection{Thermal-cycling results}

In order to study to what extent IPTLS improves thermal cycling, we 
have considered the same six symmetric TSP instances as above. 
Additionally, for comparison, we have performed thermal-cycling runs 
without IPTLS for att532 using the same local-search implementation as 
in the thermal-cycling-with-IPTLS code. The results are presented in 
Fig.\ 3 and in Table II. 

Fig.\ 3 illustrates the high efficiency of the thermal-cycling 
approach: For att532, the performance of the original algorithm 
(without IPTLS) is clearly better than that of multi-start local 
search with IPTLS for $N_{\rm a} = 1$. It is comparable to that of 
multi-start local search with IPTLS, applied to archives of 3 states, 
cf.\ Fig.\ 2. In detail, original thermal cycling is slower if low 
accuracy is desired, and better if a high accuracy has to be achieved.
However, this ranking is certainly TSP instance dependent.

The efficiency is further improved by embedding IPTLS in thermal
cycling, see Fig.\ 3. For att532, there is a gain by a factor of 2 to 
3; it slightly increases with the accuracy demanded. The comparison 
to ``parallelized'' multi-start local search with IPTLS yields a 
surprising result: For att532, when move class (d) is considered, the 
thermal-cycling-with-IPTLS code is only slightly better than that 
program, which is considerably simpler from the conceptional point of 
view (only two adjustable parameters). 

For other TSP instances however, thermal cycling with IPTLS can be 
clearly more efficient than ``parallelized'' multi-start local search 
with IPTLS, compare Table II to Table I with respect to rat783, 
pr2392, and fl3795. It is remarkable that thermal cycling with IPTLS 
reproduced the best known tour lengths for all the problems considered 
within ``reasonable'' computing times. Moreover, comparing Table II 
with Table I from \cite{thercycl} shows that, for the instances 
pcb442, att532, and rat783, the thermal-cycling-with-IPTLS program is 
typically by a factor of 2 to 3 faster than the code used in 
\cite{thercycl}. For fl1577, the acceleration amounts to a factor of 
5, and, for fl3795, it is even larger -- roughly a factor of 10 is 
obtained.

It is definitely problematic to use results obtained on different 
computing platforms (hardware, operating system, and programming 
language) as the basis of a judgement. Nevertheless, we now compare
the performance of our thermal-cycling-with-IPTLS procedure with that 
of four other approaches, but the results should be considered with 
care.

It seems that our code is more efficient, for all instances but 
pcb442, than the genetic-local-search algorithm presented in 
\cite{Merz.Frei} -- a significantly improved version of the winning 
algorithm of the {\it First International Contest on Evolutionary 
Optimization} \cite{Bers.etal,Frei.Merz.a}. In detail, our code is 
slightly slower for pcb442 and slightly faster for rat783, it has 
clear advantages for att532, in particular when high accuracies have 
to be achieved, and it is considerably faster for fl1577 and fl3795. 
However, the approach presented in \cite{Merz.Frei} has been 
optimized for solving large TSP instances by minimizing memory 
requirements; the distances between cities are computed rather than 
looked up in a distance table stored in the main memory. For example, 
the genetic-local-search algorithm of Ref.\ \cite{Merz.Frei} needs 10 
MBytes of main memory for solving fl3795 compared to 256 MBytes used 
by the program presented here.

In comparison to the iterated Lin-Kernighan approach proposed by 
Johnson and McGeoch, the performance of which is illustrated by Table 
16 of \cite{John.McGe}, our program is slower by roughly a factor of 4 
for pcb442 and att532 if the accuracy of our results for 
$N_{\rm a} = 3$ is required. The performance gap seems to shrink with 
increasing accuracy demand \cite{John}. However, for fl3795, our code 
performs considerably better. According to further calculations, this 
advantage arises primarily from a larger robustness of our code, and 
not from better scalability \cite{John}.

Clearly, one should also attempt to make a comparison with the 
state-of-the-art exact algorithms. For the Padberg-Rinaldi 532 cities 
problem, the branch-and-cut program by Thienel and Nadeff, one of the 
presently fastest exact solution codes, needs 16.5 minutes on a 
SPARC10 \cite{Thie}, which corresponds to roughly 4 minutes computing 
time for our CPU. Utilizing an archive of 12 states and cyclically 
quenching according to stage (d), we performed 100 runs. Our Monte 
Carlo approach, i.e., thermal cycling extended by IPTLS, reproduced 
the optimum tour length 27686 in 97 of the 100 runs, requiring on the
average 246 CPU seconds for one of them. In the other three cases, we 
obtained tours with lengths 27693 (once) or 27698 (twice). However, 
when comparing with exact algorithms for fl1577, the usefulness of the
proposed approach is more obvious: Here, using an archive of 8 states,
we obtained the exact optimum in 19 of 20 runs, and in one case a tour
of length 22253. Our Monte Carlo optimization requires 1390 CPU 
seconds on the average, whereas $10^6$ CPU seconds were needed for the 
only recently obtained exact solution of this problem on a DEC 
Alphastation 4100 5/400 \cite{Apple}.

Again, these comparisons should be interpreted with care: On the one 
hand, computing provably optimal solutions requires much more effort 
than simply trying to find high-quality solutions without any 
guarantee. However, on the other hand, the required effort depends not 
only on the size of the TSP instance, but also on its ``character''. 
Thus, fl3795, for which our code yields solutions of the best known 
tour length with high probability within ``reasonable'' 
$\tau_{\rm {CPU}}$, has --  to the best of our knowledge -- not been 
solved exactly yet.

\section{Conclusions}

We have presented an algorithm by means of which the effectiveness
of local search based heuristic combinatorial optimization procedures
can be increased considerably. This algorithm, iterative partial 
transcription, is physically motivated: for a spin glass, it 
corresponds to searching for non-interacting clusters of spins 
with respect to which two states differ, and relaxing these 
excitations. Mathematically spoken, the algorithm can be understood as 
a search in the subspace spanned by the differing components of two 
approximate solutions of the optimization problem. It transcribes 
subsets of the components of the vector, representing one approximate 
solution, by the related components of the other approximate solution 
if the quality of the former solution can be increased in this way. 
This process continues iteratively, accounting for an increasing 
number of components.

For the traveling salesman problem, we have demonstrated the
feasibility of this approach by embedding it in the 
multi-start-local-search algorithm starting from random states, and, 
alternatively, in the thermal-cycling method. In both cases, a 
considerable acceleration of the computation of high-quality 
approximate solutions was reached. For the TSP instances
considered, these algorithms are far more efficient than SA.

There are several areas for future research, such as (i) evaluating 
the performance of iterative partial transcription for very large TSP 
instances, (ii) investigating its usefulness in other combinatorial 
optimization problems, and (iii) incorporating it in other heuristic 
combinatorial optimization procedures.

\section*{Acknowledgements}

This work was supported by the SMWK and DFG (SFB 393). We are 
particularly indebted to M.~Pollak and U.~R\"o{\ss}ler for a 
series of critical remarks, and to R.~M\"obius for her help in finding 
an appropriate name for the procedure presented. Moreover, discussions 
with A.~D\'{\i}az-S\'anchez, D.S.~Johnson, J.~Talamantes, and 
S.~Thienel were very useful.

\begin{table}
\begin{tabular}{rrrrrrrr}
Problem&$K$&$N_{\rm a}$&$L_{\rm {min}}$&$L_{\rm {max}}$&
$n_{\rm best}$&$L_{\rm {mean}}$&$\tau_{\rm {CPU}}$\\ 
\hline
pcb442&50&1&50778&50976&1&50906&10\\
pcb442&500&1&50778&50907&15&50794&108\\
pcb442&500&10&50778&50795&18&50780&116\\ \hline
att532&50&1&27717&27803&0&27755&12\\
att532&500&1&27686&27737&8&27700&119\\
att532&500&10&27686&27693&13&27688&135\\ \hline
rat783&50&1&8823&8874&0&8849&17\\
rat783&500&1&8809&8839&0&8823&162\\
rat783&500&10&8806&8826&3&8815&191\\ \hline
fl1577&50&1&22250&22308&0&22266&113\\
fl1577&500&1&22249&22254&8&22251&1110\\
fl1577&500&10&22249&22249&20&22249&1160\\ \hline
pr2392&50&1&382178&385182&0&383675&139\\
pr2392&500&1&380776&383232&0&382255&1310\\
pr2392&500&10&379915&382980&0&381757&1790\\ \hline
fl3795&50&1&28774&28907&0&28815&781\\
fl3795&500&1&28772&28783&12&28775&7410\\
fl3795&500&10&28772&28779&13&28773&8680\\
\end{tabular}

\caption{
``Parallel'' multi-start local search with IPTLS: dependence of the 
tour length of the approximate solution, and of the computing time on 
the number of searches $K$, and on the archive size $N_{\rm a}$ for 
six instances of the TSPLIB95 \protect\cite{Rein}, where the local 
search algorithm is based on move class (d). For series of 20 runs, 
smallest and largest tour lengths, $L_{\rm {min}}$ and 
$L_{\rm {max}}$, number of obtaining the best known tour length,
$n_{\rm best}$, mean tour length, $L_{\rm {mean}}$, and computing time 
in seconds, $\tau_{\rm {CPU}}$, are given. 
}
\end{table}

\begin{table}
\begin{tabular}{rrrrrrr}
Problem&$N_{\rm a}$&$L_{\rm {min}}$&$L_{\rm {max}}$&
$n_{\rm best}$&$L_{\rm {mean}}$&$\tau_{\rm {CPU}}$\\ 
\hline
pcb442&1&50778&51024&10&50860&14\\
pcb442&3&50778&50912&15&50800&33\\
pcb442&5&50778&50912&19&50785&60\\ \hline
att532&1&27686&27742&1&27714&19\\
att532&3&27686&27718&4&27701&47\\
att532&5&27686&27704&16&27688&88\\ \hline
rat783&1&8806&8839&2&8816&29\\
rat783&3&8806&8812&6&8808&62\\
rat783&5&8806&8809&14&8806.6&112\\ \hline
fl1577&1&22249&22262&2&22255&193\\
fl1577&3&22249&22261&9&22252&450\\
fl1577&5&22249&22253&16&22249.8&826\\ \hline
pr2392&1&378579&381023&0&380036&309\\
pr2392&3&378143&379649&0&378950&927\\ 
pr2392&5&378032&379398&1&378558&2140\\
pr2392&8&378032&379000&1&378428&4700\\
pr2392&12&378032&378655&7&378158&9380\\ \hline
fl3795&1&28772&28774&19&28772.1&1520\\ 
fl3795&3&28772&28785&14&28774&3110\\ 
fl3795&5&28772&28772&20&28772&6050\\ 
\end{tabular}

\caption{
Thermal cycling \protect\cite{thercycl} extended by IPTLS: 
dependence of tour length, number of obtaining the best known tour
length, and computing time on the archive size. For details see 
caption of Table I.
}
\end{table}

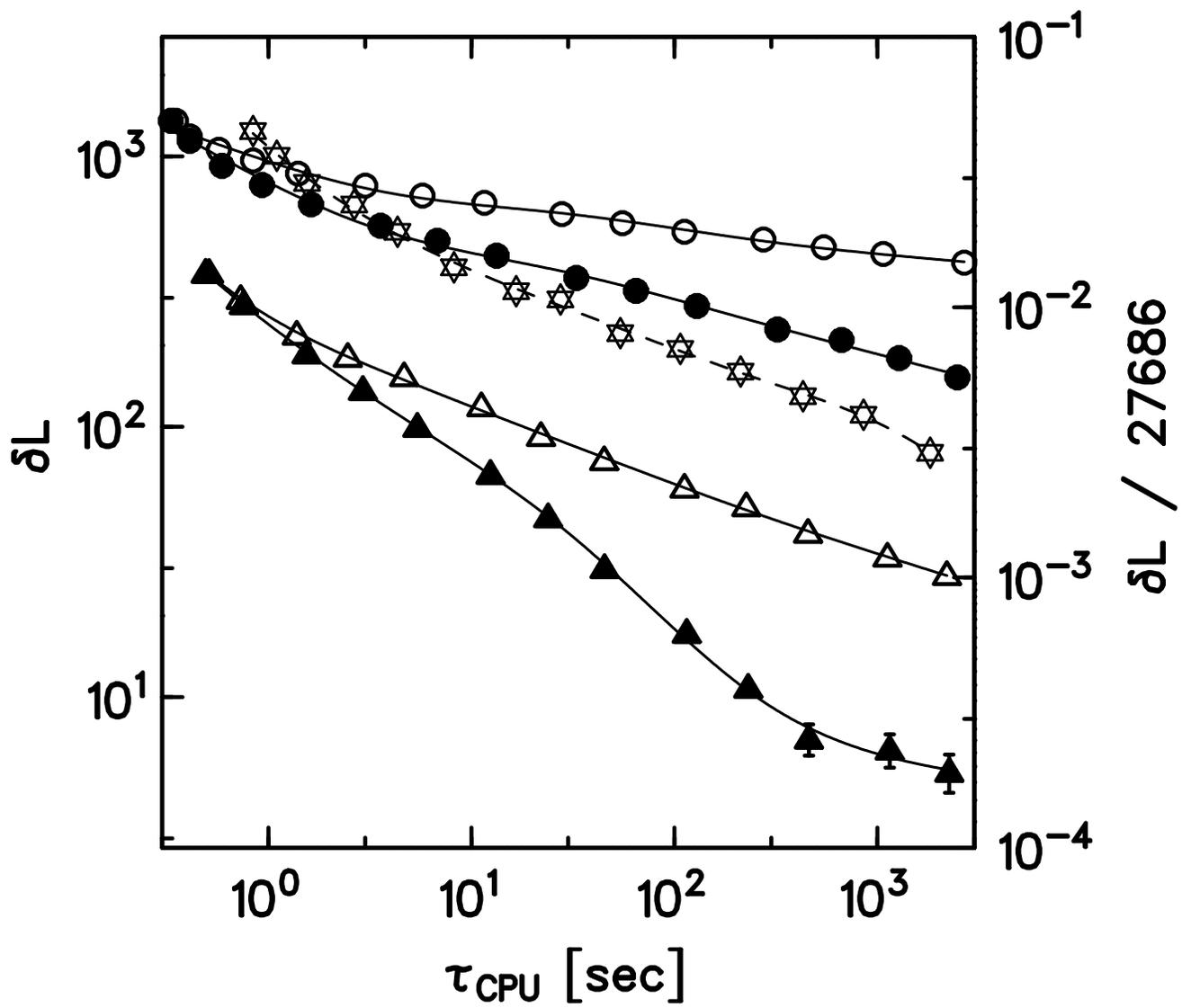
\begin{figure}
\caption{
Effect of embedding IPTLS in multi-start local search: relation 
between computing time $\tau_{\rm {CPU}}$ (in seconds) and average 
deviation, $\delta L = L_{\rm {mean}} - 27686$, of the obtained 
approximate solution from the optimum tour length for the 
Padberg-Rinaldi 532 cities problem, att532. $\circ$ ($\bullet$) and 
$\triangle$ ($\blacktriangle$): multi-start local search based on 
move classes (a) and (d), respectively, without (with) IPTLS; 
${\scriptscriptstyle ^{\textstyle \vartriangle}} \!\!\!\! 
{\scriptscriptstyle \!\!\:} \triangledown$: SA. In all cases, averages 
were taken from 100 runs; fluctuations ($1\sigma$-region) are indicated 
by error bars if they exceed the symbol size. The lines, full for 
multi-start local search, and dashed for SA, are guides to the eye only.
}
\label{fig1}
\end{figure}

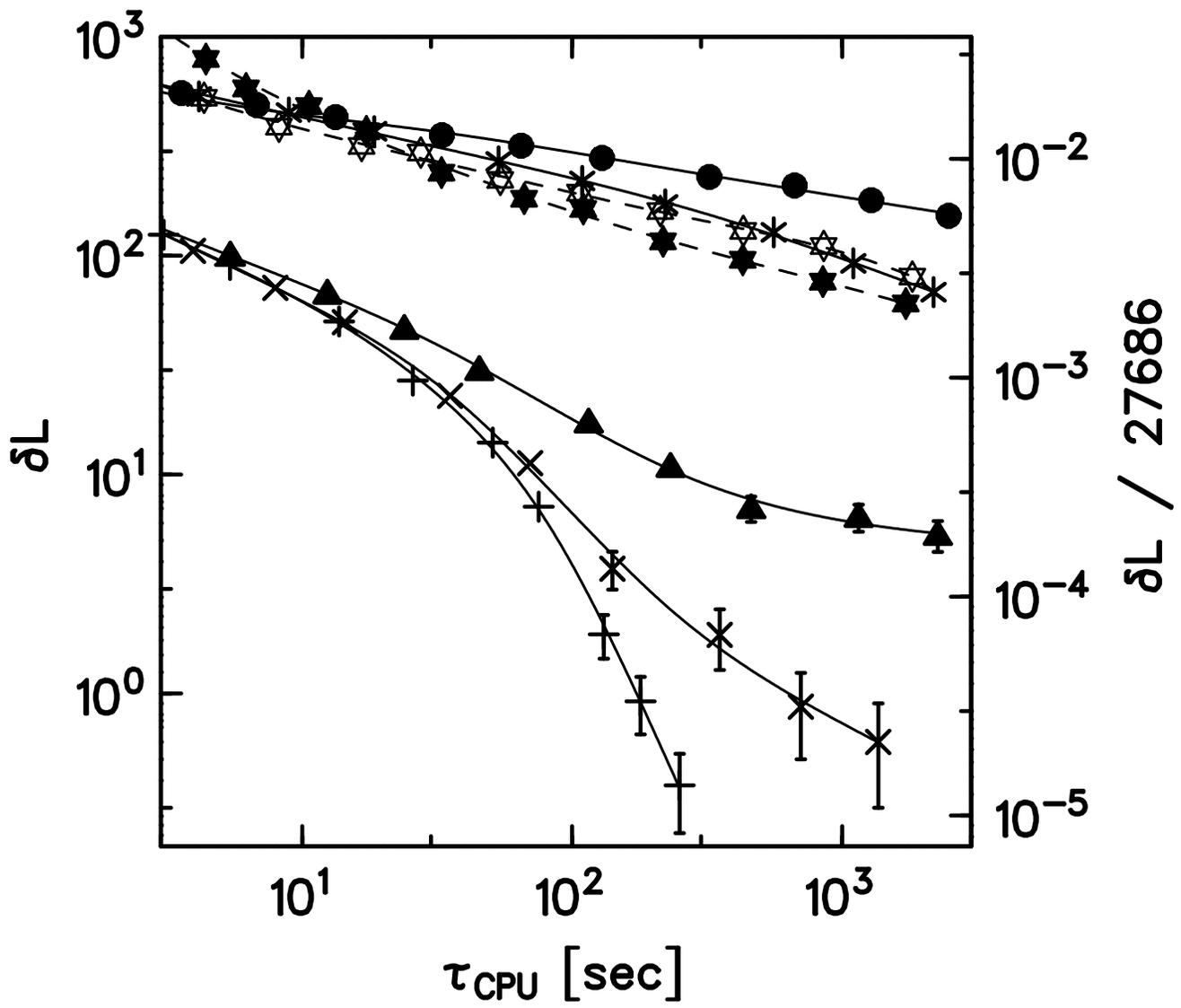
\begin{figure}
\caption{
Effect of ``parallelizing'' multi-start local search with IPTLS:
mean deviation from optimum tour length in dependence on computing 
time for att532. $\bullet$ and {\large $\ast$}: $N_{\rm a} = 1$ and 10, 
respectively, for move class (a); $\blacktriangle$, $\times$, and 
+: $N_{\rm a} = 1$, 3, and 10, respectively, for move class (d); 
${\scriptscriptstyle ^{\textstyle \vartriangle}} \!\!\!\! 
{\scriptscriptstyle \!\!\:} \triangledown$ and ${\scriptscriptstyle 
^{\textstyle \blacktriangle}} \!\!\!\! {\scriptscriptstyle \!\!\:} 
\blacktriangledown$: SA without and with ``parallelizing'', 
respectively. For further details see caption of Fig.\ 1.
}
\label{fig2}
\end{figure}

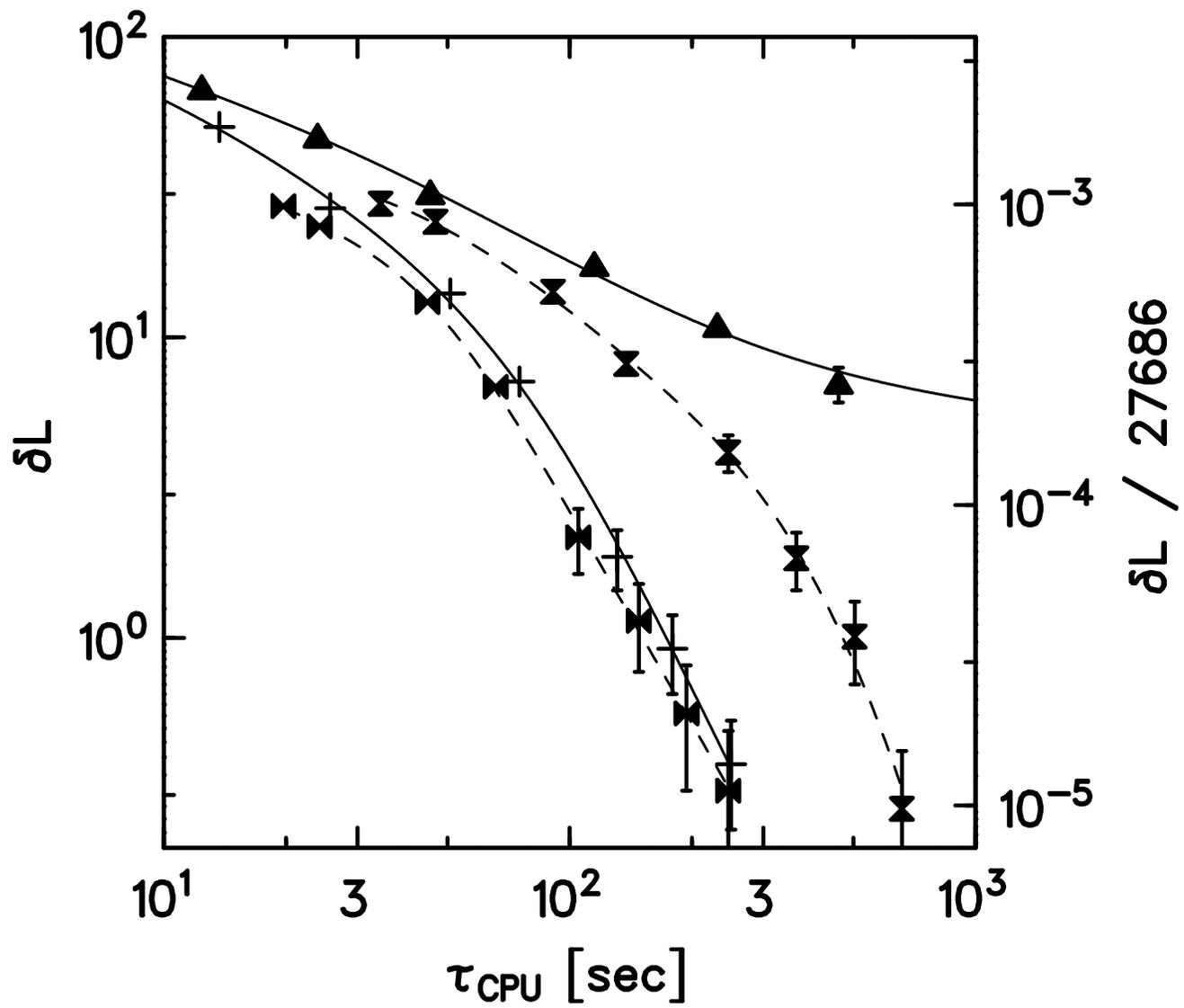
\begin{figure}
\caption{
Effect of embedding IPTLS in thermal cycling: mean deviation from 
optimum tour length in dependence on computing time for att532. 
${\scriptstyle \blacktriangle \!\! {\scriptscriptstyle \!}} \! 
^{\scriptstyle \blacktriangledown}$ (${\scriptstyle \blacktriangleright 
{\scriptscriptstyle \!} \! \blacktriangleleft}$): thermal cycling 
without (with) IPTLS; $\blacktriangle$ (+): multi-start local search 
with IPTLS for $N_{\rm a} = 1$ (10), included for comparison. In all 
cases, move class (d) is considered. The lines, full for multi-start 
local search, and dashed for thermal cycling, are guides to the eye 
only. For further details see caption of Fig.\ 1.
}
\label{fig3}
\end{figure}


\begin{references}
\bibitem{touch} The comparison needs an effort only proportional to $s$ 
   when touch bits are ascribed to the cities.
\end{references}
\end{document}